\documentclass[twocolumn,showpacs,amssymb,aps,pra,amsmath,superscriptaddress,reprint]{revtex4-1}

\usepackage[colorlinks,pdfusetitle,urlcolor=blue,citecolor=blue,linkcolor=blue,bookmarksnumbered,plainpages=false]{hyperref}

\usepackage{graphicx}
\usepackage{dcolumn}
\usepackage{bm}
\usepackage{mathtools}
\usepackage{nicefrac}
\usepackage{cancel}
\usepackage[T1]{fontenc}
\usepackage{textcomp}
\usepackage{gensymb}

\usepackage{color}

\newcommand{\B}[1]{\mathbf{#1}}

\renewcommand{\vec}[1]{\bm{#1}}
\newcommand{\tens}[1]{\mbox{\textbf{{\textsf{#1}}}}}

\newcommand{\mi}{\mathrm{i}}

\newcommand\elec[2]{\;
\begin{matrix} #1 \\ #2 \\ \end{matrix}\;}
\newcommand{\Tr}{\,\mathrm{Tr}\,}
\newcommand{\inte}[3]{\int_{#1}^{#2}\!\!\!\!\! \mbox{d}#3\;}

\begin{document}

\title{Impact of Casimir-Polder interaction on Poisson-spot diffraction at a dielectric sphere }

\author{Joshua L. Hemmerich}
\affiliation{Physikalisches Institut, Albert-Ludwigs-Universit\"{a}t Freiburg, Hermann-Herder-Str. 4, D-79104, Freiburg i. Br., Germany}
 \author{Robert Bennett}
 \affiliation{Physikalisches Institut, Albert-Ludwigs-Universit\"{a}t Freiburg, Hermann-Herder-Str. 4, D-79104, Freiburg i. Br., Germany}
  \author{Thomas Reisinger}
  \affiliation{Institute of Nanotechnology, Karlsruhe Institute of Technology, Hermann-von-Helmholtz-Platz 1, D-76344, Karlsruhe, Germany}
    \author{Stefan Nimmrichter}
 \affiliation{Faculty of Physics, University of Duisburg-Essen, Lotharstr. 1, 47048 Duisburg, Germany}
  \author{Johannes Fiedler}
 \affiliation{Physikalisches Institut, Albert-Ludwigs-Universit\"{a}t Freiburg, Hermann-Herder-Str. 4, D-79104, Freiburg i. Br., Germany}
  \author{Horst Hahn}
  \affiliation{Institute of Nanotechnology, Karlsruhe Institute of Technology, Hermann-von-Helmholtz-Platz 1, D-76344, Karlsruhe, Germany}
  \author{Herbert Gleiter}
    \affiliation{Institute of Nanotechnology, Karlsruhe Institute of Technology, Hermann-von-Helmholtz-Platz 1, D-76344, Karlsruhe, Germany}
 \author{Stefan Yoshi Buhmann}
 \affiliation{Physikalisches Institut, Albert-Ludwigs-Universit\"{a}t Freiburg, Hermann-Herder-Str. 4, D-79104, Freiburg i. Br., Germany}
 \affiliation{Freiburg Institute for Advanced Studies, Albert-Ludwigs-Universit\"{a}t Freiburg, Albertstr. 19, D-79104 Freiburg i. Br., Germany}

\date{\today}
\begin{abstract}
Diffraction of matter-waves is an important demonstration of the fact that objects in nature possess a mixture of particle-like and wave-like properties. Unlike in the case of light diffraction, matter-waves are subject to a vacuum-mediated interaction with diffraction obstacles. Here we present a detailed account of this effect through the calculation of the attractive Casimir-Polder potential between a dielectric sphere and an atomic beam. Furthermore, we use our calculated potential to make predictions about the diffraction patterns to be observed in an ongoing experiment where a beam of indium atoms is diffracted around a silicon dioxide sphere. The result is an amplification of the on-axis bright feature which is the matter-wave analogue of the well-known `Poisson spot' from optics. Our treatment confirms that the diffraction patterns resulting from our complete account of the sphere Casimir-Polder potential are indistinguishable from those found via a large-sphere non-retarded approximation in the discussed experiments, establishing the latter as an adequate model. 
\end{abstract}


\maketitle

\section{\label{IntroSection} Introduction}

Matter-wave diffraction around material objects is one of the most compelling demonstrations of the particle-wave duality. Beginning from the classic electron diffraction experiments of the 1920s \cite{Davisson1927,Thomson1927}, particles of progressively higher mass have had their wave-like nature revealed. This process began in the 1930s and 1940s with the diffraction of atoms and molecules \cite{stern1930} as well as neutrons \cite{Shull1948} from various crystal surfaces. More recently the diffraction of atoms \cite{Keith1988} and simple molecules \cite{Schollkopf1994} from lithographically fabricated grating structures have been demonstrated. These experiments paved the way for grating diffraction experiments with complex organic molecules such as fullerenes \cite{Arndt1999} and porphyrin derivates \cite{Gerlich2011}. Scaling of diffraction experiments to even larger objects such as macromolecules or even living organisms like viruses or bacteria presents some considerable difficulties, but has the potential to shed light on the question if quantum mechanics applies unmodified to such increasingly macroscopic systems \cite{bassi_models_2013}. The latter used a Talbot-Lau arrangement of three gratings, with the middle grating realized by a standing light-wave to eliminate the problem of molecule-grating interaction.

As the diffracting molecules become larger, a number of difficulties arise. The immediate reduction in de-Broglie wavelength, can be counteracted for example by a reduction in the molecular speeds or the use of gratings with smaller grating constants, which both present significant technological challenges. In addition, interaction with the environment, for example via thermal emission of radiation \cite{hackermueller2004}, can lead to decoherence \cite{hornberger2004}. More practically, the buildup of unwanted contaminants (from the beam or elsewhere) upon the grating itself over time can result in a reduction in the interference visibility or even cause the slits to become blocked. Additionally, Talbot-Lau interferometers impose relatively loose restrictions on the width of the beam's wavelength distribution, which may however become limiting for sources of objects of increasing mass such as, for example, cluster sources.  Furthermore, the gratings must have very uniform grating constants and must be aligned with high precision. 

\begin{figure}[t!]
\centering
\includegraphics[width=0.75\columnwidth]{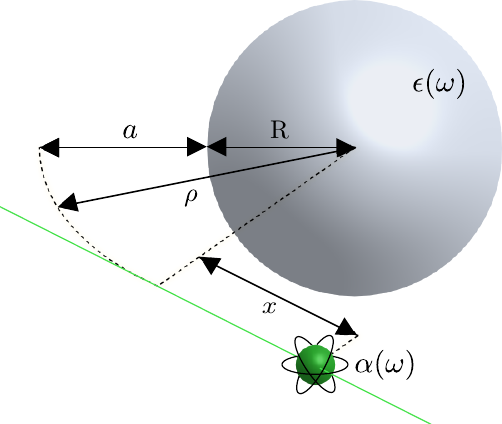}
\caption{Schematic representation of the system we consider, where an atom of polarisability $\alpha(\omega)$ passes near a sphere of permittivity $\epsilon(\omega)$. The resulting Casimir-Polder/van der Waals interaction will cause the atomic wave function to pick up a phase shift, which is observable in matter-wave diffraction experiments.}
\label{SphereDiagram}
\end{figure}

Finally, the problem addressed in the present study stems from the fact that extended particles undergoing diffraction have a non-zero electromagnetic polarisability in general, meaning they experience Casimir-Polder/Van der Waals (CP/vdW) dispersion forces originating from the grating itself. The result is an effective reduction in the slit width in addition to a coherent phase shift \cite{Perreault2005b}, which increasingly obscures the distinction between particle and wave nature \cite{hornberger2004,reisinger_particlewave_2011}. The current mass record is held by a setup that reduces this problem through the use of a standing light-wave phase-grating as the middle diffraction grating in a Talbot-Lau interferometer arrangement \cite{eibenberger_matterwave_2013}. Another approach has the potential to eliminate the problem entirely, by using three pulsed laser-ionization gratings \cite{haslinger_universal_2013}. Aside from these developments, an accurate knowledge of the CP/vdW incurred phase-shifts is highly desirable, but remains challenging. The dispersion forces can exhibit intricate spatial dependence in complex geometries (see, for example, \cite{Henkel1998,Messina2009,Contreras-Reyes2010,Bennett2015,Rodriguez2007,Eberlein2011}), but in many far-field diffraction experiments the effect is reduced to an effective slit-narrowing fitted to the data after the experiment \cite{Nairz2003}. Dispersion forces near gratings are extremely difficult to model accurately \cite{Eberlein2011,Henkel1998,Messina2009,Rodriguez2007,Contreras-Reyes2010,Bennett2015,Grisenti1999a,Grisenti2000,Brevik1998,Cronin2004,Perreault2005b,Cronin2005,Neto2005a,Bimonte2014,Bender2014}, especially when sharp edges are involved \cite{Gies2006}. This, coupled to the fact that gratings necessarily have a large number of sharp edges spaced closely together, means that progress in detailed accounts of this effect has stalled. 

Here we investigate a different type of diffraction scheme --- the `Poisson spot' interferometer. There, waves in general are diffracted around a circular or spherical object, resulting in an on-axis bright spot, called Poisson's spot or spot of Arago \cite{harvey_spot_1984,hecht}. This effect was first predicted by Poisson when looking for evidence against Fresnel's wave theory of light in the early 1800s. Poisson described it as an absurd prediction of Fresnel's theory, but experiments by Fran\c{c}ois Arago proved that the effect is real, accelerating the shift away from Newton's `corpuscular' theory \cite{gooding1989uses}. The matter-wave version of this experiment \cite{Reisinger2009} avoids some of the problems that appear in the grating experiments discussed above (e.g. blocking of the grating, or alignment). However, the inevitable contribution from CP/vdW forces remains. These have been accounted for using a relatively simple model for the disk-based experiments of \cite{Reisinger2009} in \cite{reisinger_particlewave_2011}. The question of whether this approach is valid is part of the motivation for the study presented here.

The current article covers an approach related to \cite{Reisinger2009}, where a dielectric sphere [see Fig. \ref{SphereDiagram}] replaces the discs used as diffraction obstacles in the aforementioned studies. The lack of sharp edges in spherical diffraction obstacles makes this approach ideally suited to accurate modelling of the CP/vdW interaction, see, for example \cite{Juffmann2012b}. The focus is in particular directed at a specific experiment that is currently being conducted at the Karlsruhe Institute of Technology, the parameters of which will be used throughout this article. The experiment aims at recording diffraction patterns in the shadow of sub-micron sized silica particles cast by thermally evaporated indium atoms. The choice of these two materials is motivated in the following ways. 

The St\"{o}ber process \cite{stober_controlled_1968} enables the controlled growth of monodispersed spheres composed of amorphous silicon-dioxide via condensation of silicic acid in alcoholic solutions. The advantage in preparing these silica spheres, which are to be used as diffraction obstacles, using a bottom-up approach as opposed to a top-down lithographic process is the close to optimal spherical shape that can be achieved in this way. Specifically the low surface corrugation of the particles is crucial \cite{Reisinger2009} for achieving the Poisson spot visibilities reported here. Furthermore, the resulting huge quantities of diffraction obstacles of uniform size are compatible with a simple parallelization of the experiment in order to average over large numbers of recorded diffraction images. 

The choice of indium is largely due to its high vapor pressure, which enables the realization of a thermal-oven-based point-source of sufficient brightness. It also sets the experiment apart from matter-wave diffraction aimed at measuring CP/vdW forces with beams composed of alkali metal atoms \cite{Perreault2005a,Perreault2005b}, and seeks to demonstrate in this way a compatibility with a large number of condensable atom and molecular species. Thus, a wide variety of CP/vdW potentials could be studied using the same experimental approach.

In the following section we review the derivation of CP/vdW potentials and apply it to the particular materials and geometry used in the experiment. Then, in section \ref{MatterWaveScatteringSection} we derive the resulting phase shifts affecting the matter-wave diffraction experiment. Finally, a numerical solution of the Fresnel diffraction integral is used in section \ref{ExperimentSection} to predict the Poisson spot diffraction intensities in the presence of the CP/vdW potential followed by a discussion and conclusion.

\section{Casimir-Polder Potential}\label{CPPotentialSection} 
In this section we outline a general derivation for the CP interaction for a dielectric sphere and a single ground-state atom and show that it reduces to well-known results in asymptotic cases. 
\subsection{Perturbation theory}

We consider an atomic dipole $\hat{\B{d}}$ interacting with the electric field $\hat{\B{E}}$ of macroscopic QED \cite{Gruner1996a, Scheel2009}, which includes all the information about geometry and dielectric functions in the electromagnetic environment surrounding the atom. 
We will work in the long-wavelength approximation, where one can restrict to the first term in the multipole expansion of the atom-field interaction, meaning that the interaction Hamiltonian is:
\begin{equation}
H = -\hat{\B{d}}\cdot\hat{\B{E}}(\B{r}_\text{A})
\end{equation}
with 
\begin{equation}\label{mqedEField}
\hat{\B{E}}(\B{r})=\int d^3 \B{r}'\int_0^\infty d\omega\, \tens{G}(\B{r},\B{r}',\omega)\cdot \hat{\B{f}} (\B{r}',\omega)+\text{h.c.}
\end{equation}
where $\hat{\B{f}} (\B{r}',\omega)$ is a bosonic field operator that describes the fundamental excitations of the composite matter-field system and $\tens{G}(\B{r},\B{r}',\omega)$ is the electromagnetic Green's tensor solving
\begin{equation}
\nabla \times \nabla \times \tens{G}(\B{r},\B{r}',\omega) - \varepsilon(\B{r},\omega) \frac{\omega^2}{c^2}\tens{G}(\B{r},\B{r}',\omega) = \delta(\B{r}-\B{r}')
\end{equation}
where $\varepsilon(\B{r},\omega)$ is the position and frequency-dependent permittivity of the system. Application of second-order perturbation theory to the ground state of the atom-field system yields the following expression for the CP potential in terms of a complex frequency $\omega = \mi\xi$ \cite{Buhmann2004,buhmann2013dispersion} 
\begin{equation}
\label{CPpotential}
U(\vec{r})=\frac{\hbar\mu_0}{2\pi}
\inte{0}{\infty}{\xi}\,\xi^2\alpha(\mi\xi)\Tr
\tens{G}^{(1)}(\vec{r},\vec{r},\mi\xi)\;.
\end{equation}
where $\tens{G}^{(1)}(\vec{r},\vec{r}',\mi\xi)$ is the scattering part of the Green's tensor for the geometry at hand, which is obtained from the full Green's tensor at each point by subtracting the Green's tensor of a homogenous material with the same permittivity as the point in question. In this work we are only interested in the case when the atom is in the vacuum region outside a sphere, so for our purposes the scattering Green's tensor is obtained simply by subtracting the vacuum Green's tensor from that of a sphere. The quantity $\alpha(\omega)$ in Eq.~\eqref{CPpotential} is the atomic polarizability for a transition from the ground state to state $k$ with dipole moment $d_{0k}$ and frequency $\omega_{0k}$;
\begin{equation}
\label{pol}
\alpha(\omega) =\frac{2}{3\hbar(2J_0+1)}\sum_{k}\frac{\omega_{0k}d^2_{0k} }{\omega_{0k}^2-\omega^2}
\end{equation}
where $J_0$ is the total angular momentum quantum number of the ground state, which here appears as a weighting factor accounting for its degenerate levels. 

The Green's tensor outside a sphere can be written as \cite{Le-WeiLi1994,tai1994dyadic,buhmann2013dispersion} \begin{multline}
\tens{G}^{(1)}(\mathbf{r},\mathbf{r}',\omega)=\frac{\mathrm{i}k}{4\pi}\sum\limits_{l=1}^{\infty}\sum\limits_{m=0}^{l}\sum\limits_{\sigma=\mbox{\tiny TE,TM}}(2-\delta_{m0}) \\\times\frac{2l+1}{l(l+1)}\frac{(l-m)!}{(l+m)!}\: r_{l\sigma}\big[\mathbf{a}_{lm+\sigma}(\mathbf{r})\otimes\mathbf{a}_{lm+\sigma}(\mathbf{r}') \\+\mathbf{a}_{lm-\sigma}(\mathbf{r})\otimes\mathbf{a}_{lm-\sigma}(\mathbf{r}')\big]\: .
\label{tens_sphere}
\end{multline}
where $\otimes$ denotes the dyadic product $(\mathbf{A}\otimes \mathbf{B})_{ij}=A_iB_j$ and $\mathbf{a}_{lm\pm\sigma}$ are spherical wave-vector functions as listed in Appendix \ref{Appendix}, and $r_{l\sigma}$ are the Mie reflection coefficients for a sphere of radius $R$, given by; 
\begin{multline}
\label{mie_refl}
r_{lTE}=-\frac{j_l(z_1)[z_2j_l(z_2)]'-[z_1j_l(z_1)]'j_l(z_2)}{h^{(1)}_l(z_1)[z_2j_l(z_2)]'-[z_1h^{(1)}_l(z_1)]'j_l(z_2)}\:, \\
r_{lTM}=-\frac{\varepsilon(\mathrm{i}\xi)j_l(z_2)[z_1j_l(z_1)]'-j_l(z_1)[z_2j_l(z_2)]'}{\varepsilon(\mathrm{i}\xi)[z_1h^{(1)}_l(z_1)]'j_l(z_2)-h^{(1)}_l(z_1)[z_2j_l(z_2)]'}\:,
\end{multline}
where $z_i\equiv k_iR$, $k_1=\omega/c$, $k_2=\omega\sqrt{\varepsilon(\omega)}/c$ and the primes denote derivatives with respect to $z_1$, $z_2$ respectively, e.g. $[x f(x)]' \equiv \frac{d}{dx} [xf(x)]$. The quantities $j_l(z_i)$ and $h_l^{(1)}(z_i)$ are the spherical Bessel and Hankel functions of the first kind, respectively.

Using addition theorems  for spherical harmonics \cite{jackson1998classical}  , the potential for a sphere can be rewritten in the following form \cite{Buhmann2004b} (see Appendix \ref{GTAppendix}):
\begin{multline}
U(r)=-\frac{\hbar\mu_0}{8\pi^2c}\int\limits_0^\infty d\xi\xi^3\alpha(\mathrm{i}\xi)\sum\limits_{l=1}^{\infty}(2l+1) \\  \times\bigg\{r_{lTE}[h^{(1)}_l(kr)]^2
+r_{lTM}\bigg[l(l+1)\frac{[h^{(1)}_l(kr)]^2}{(kr)^2} \\+\frac{[krh^{(1)}_l(kr)]'^2}{(kr)^2}\bigg]
\bigg\}.
\label{cp_potential_sphere}
\end{multline}
in agreement with \cite{Jhe1995,Jhe1995a}. We now investigate Eq.~(\ref{cp_potential_sphere}) in several asymptotic regimes, both as a consistency check and as a useful point of comparison later on. Firstly we consider atom-sphere distances $r-R$ much smaller than the atomic transition wavelengths $\lambda_\text{A}=2\pi c/\omega_\text{A}$, which means that the atom-sphere interaction can be considered to be instantaneous so is termed the non-retarded regime. This renders the non retarded CP potential $U_{\mbox{\tiny NR}}(r) \equiv U(r\ll \lambda_\text{A})$\cite{Jhe1995,Jhe1995a,buhmann2013dispersion}
\begin{multline}
\label{nr}
U_{\mbox{\tiny NR}}(r)=-\frac{\hbar}{8\pi^2\varepsilon_0}\sum\limits^\infty_{l=1}(2l+1)(l+1)\\
\times\frac{R^{2l+1}}{r^{2l+4}}\inte{0}{\infty}{\xi}\alpha(\mathrm{i}\xi) \frac{\varepsilon(\mathrm{i}\xi)-1}{\varepsilon(\mathrm{i}\xi)+[(l+1)/l]}.
\end{multline}
It is worth noting that this expression does not depend on the spherical Bessel and Hankel functions $j_l$ and $h^{(1)}_l$, which means the convergence of the sum over $l$ is much more robust than for the full potential \eqref{cp_potential_sphere}. Considering furthermore the sphere radius $R$ to be much greater than the distance $r$ from the surface of the sphere to the atom  $|R-r|=z \ll R$, the terms with large $l$ dominate and yield a $1/z^3$ dependence 
\begin{equation}
\label{bnr}
U_{\mbox{\tiny BNR}}=
-\frac{\hbar }{16\pi^2\varepsilon_0}\frac{1}{z^3}\inte{0}{\infty}{\xi}\alpha(\mathrm{i}\xi)\frac{\varepsilon(\mathrm{i}\xi)-1}{\varepsilon(\mathrm{i}\xi)+1} \equiv -\frac{C_3}{z^3}
\end{equation}
where the coefficient $C_3$ has been defined for later use. Equation \eqref{bnr} is the well-known Lennard-Jones formula \cite{Lennard-Jones1932} for the potential near a half space, which is indeed the expected limiting case for an atom a small distance from a large sphere. 

Separately, we can consider a small sphere radius compared to the distance to the atom $r \gg R$. Beginning again from the general formula \eqref{cp_potential_sphere}, one finds the leading-order term in this expansion for small  $R/r$  comes from the first spherical harmonic $l=1$, which leads to the small-sphere potential:
\begin{multline}
U_{\mbox{\tiny S}}(r)=-\frac{\hbar}{4\pi^2\varepsilon_0}\frac{R^3}{r^6}\inte{0}{\infty}{\xi}\alpha(\mathrm{i}\xi)\frac{\varepsilon(\mathrm{i}\xi)-1}{\varepsilon(\mathrm{i}\xi)+2}\exp(-2\xi r/c)\\
\times\left[3+6(\xi r/c)+5(\xi r/c)^2+2(\xi r/c)^3+(\xi r/c)^4\right].
\end{multline}
which can again be further simplified in the non-retarded regime $U_{\mbox{\tiny SNR}}(r) \equiv U_{\mbox{\tiny S}}(r\ll \lambda)$
\begin{equation}
\label{snr}
U_{\mbox{\tiny SNR}}(r)=-\frac{3\hbar c}{4\pi^2\varepsilon_0}\frac{R^3}{r^6}\inte{0}{\infty}{\xi}\alpha(\mathrm{i}\xi)\frac{\varepsilon(\mathrm{i}\xi)-1}{\varepsilon(\mathrm{i}\xi)+2}
\end{equation}
which is the well-known Van der Waals potential between two microscopic polarisable objects \cite{Eisenschitz1930}. 

Finally, considering the limit of strong retardation of the electromagnetic field $\lambda_\text{A}\ll r$ and a point-like sphere as before, the retarded CP potential is obtained \cite{Casimir1948a,Nabutovskii1979,Jhe1995,Jhe1995a}
\begin{equation}
\label{sr}
U_{\mbox{\tiny SR}}(r)=-\frac{23\hbar c}{16\pi^2\varepsilon_0}\frac{R^3}{r^7}\alpha(0)\frac{\varepsilon(0)-1}{\varepsilon(0)+2}
\end{equation}

So far our considerations have been for a sphere of unspecified permittivity, and a general (ground state) atom. In order to calculate this potential explicitly, one needs to use particular values of the atomic polarizability and the permittivity for the sphere as functions of imaginary frequency $\xi$. 

\section{Numerical investigation of a real system}\label{MatterWaveScatteringSection}

\subsection{Material Response functions}
As mentioned in the introduction, we specifically consider amorphous silicon dioxide (SiO$_2$) for the sphere, and indium for the atom. To determine the permittivity $\varepsilon(\mathrm{i}\xi)$ of (SiO$_2$) as a function of imaginary frequencies $\rm{i}\xi$ we used tabulated data for the real-frequency refractive index \cite{palik1998handbook}, which was then converted to that for imaginary frequencies via the Kramers-Kronig relations (see, for example, \cite{jackson1998classical}). Amorphous SiO$_2$ has two main groups of resonances at frequencies $\omega_{T,1}$ and $\omega_{T,2}$, to which we fitted a two-line Drude model on the imaginary frequency axis:
\begin{equation}
\label{permit}
\varepsilon(\mathrm{i}\xi)=1+\frac{\omega_{P,1}^2}{\omega_{T,1}^2 + \gamma_1\xi + \xi^2}+\frac{\omega_{P,2}^2}{\omega_{T,2}^2 + \gamma_2\xi + \xi^2}\:
\end{equation}
where $\omega_{P,i}$ are the plasma frequencies of the two effective resonances and $\gamma_i$ their decay width. The explicit values of the parameters of our fit are shown in  Tab.~\ref{TAB:fit}. 
\begin{table}
\centering
\def\arraystretch{1.4}
\begin{tabular*}{\linewidth}{c @{\extracolsep{\fill}} cc|ccc}
\hline\hline
Par. & Value & Err. & Par. & Value & Err. \\
\hline
$\omega_{P,1}$ & $1.75\times10^{14}$ Hz & $0.37 \%$ & $\omega_{P,2}$ & $2.96\times10^{16}$ Hz & $0.45 \%$ \\
$\omega_{T,1}$ & $1.32\times10^{14}$ Hz & $0.34 \%$ & $\omega_{T,2}$ & $2.72\times10^{16}$ Hz & $0.43 \%$ \\
$\gamma_1$ & $4.28\times10^{13}$ Hz & $2.07 \%$ & $\gamma_2$ & $8.09\times10^{15}$ Hz & $3.40 \%$ \\
\hline\hline
\end{tabular*}
\caption{Fit parameters for a 2-line Drude-Lorentz model [Eq.~\eqref{permit}] for the permittivity of SiO$_2$.}
\label{TAB:fit}
\end{table}
The properties of the atom enter into the CP potential \eqref{cp_potential_sphere} via the polarizability \eqref{pol}, which in turn depends on the dipole matrix elements $d_{0k}$ and frequencies $\omega_{0k}$ describing transitions from the ground state to level $k$. These parameters (obtained from  \cite{Safronova2013},  where they are found by combining experimental data and computational chemistry) are listed in Tab.~\eqref{data_D} for each possible transition.
\begin{table}
\centering
\def\arraystretch{1.2}
\begin{tabular*}{\linewidth}{c @{\extracolsep{\fill}} cllll}
\hline\hline
&$k$ &  Transition & $\omega_{0k}$   & $D_{0k}$ &  \\
\hline & 1 & $5P_{1/2}\to6S_{1/2}$ & 4.594 & 16.092   &\\
	   & 2 & $5P_{1/2}\to5D_{3/2}$ & 6.200 & 22.048   &\\
       & 3 & $5P_{1/2}\to7S_{1/2}$ & 6.843 &  4.587   &\\
       & 4 & $5P_{1/2}\to6D_{3/2}$ & 7.360 &  7.910   &\\
       & 5 & $5P_{1/2}\to8S_{1/2}$ & 7.659 &  2.518   &\\
       & 6 & $5P_{1/2}\to7S_{3/2}$ & 7.886 &  3.582   &\\
\hline\hline
\end{tabular*}
\caption{Transition frequencies $\omega_{0k}$ in $10^{15}$ rad/s and dipole matrix elements $D_{0k}$ in $10^{-30}$ C$\cdot$m for indium \cite{Safronova2013}. Each $k$ represents one degenerate manifold of internal states. .}
\label{data_D}
\end{table}
The polarizability $\alpha(\mathrm{i}\xi)$ for indium in its ground state  ($5P_{1/2}$, i.e. $J_0 =1/2$ )  was then calculated from the data from Tab.~\ref{data_D} via \eqref{pol}.

\subsection{Numerical Calculations}

\begin{figure}[t]
\centering
\includegraphics[width=\linewidth]{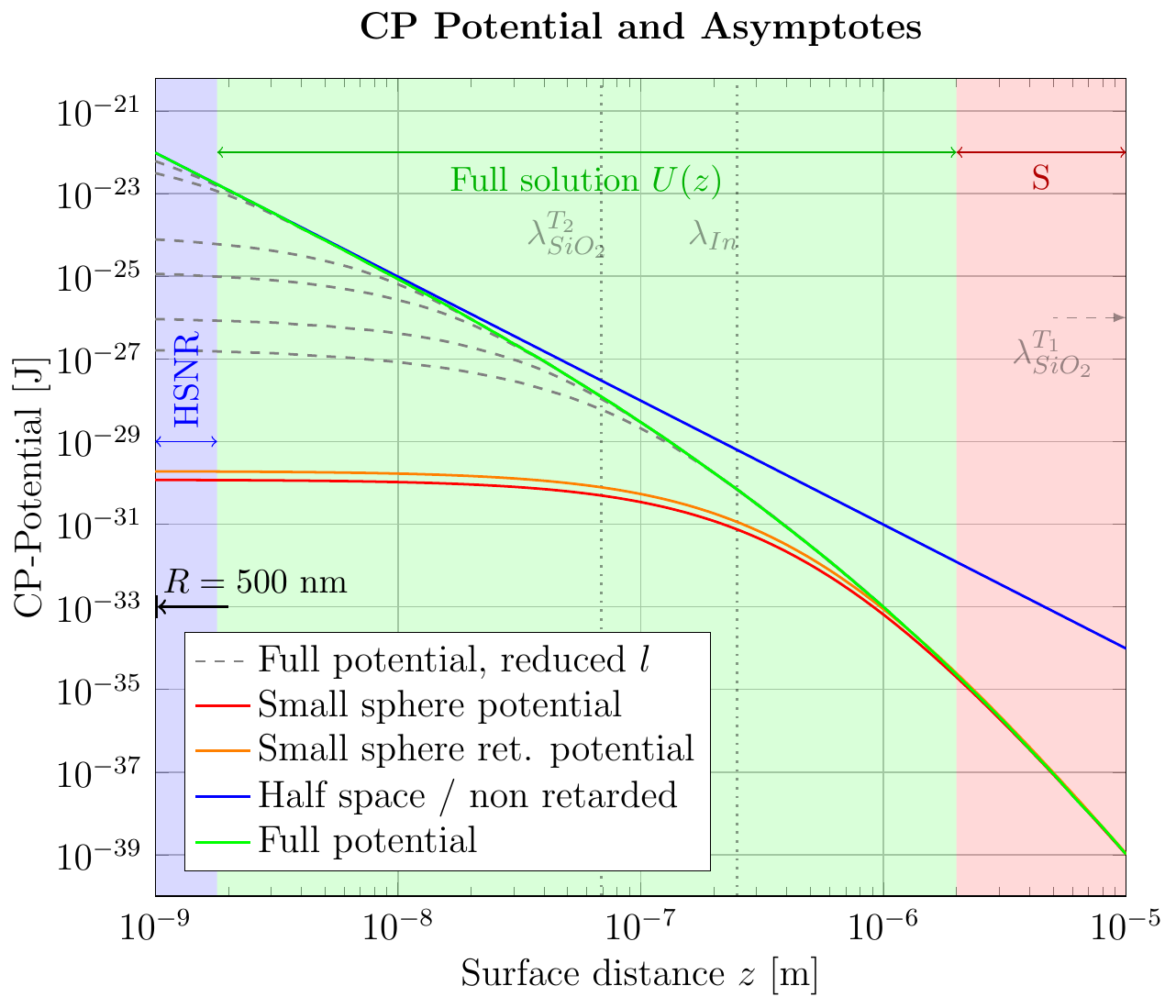}
\caption{CP potential for a $500$ nm SiO$_2$ (amorphous) sphere and a ground-state indium atom. The gray dashed lines are the solutions from the full potential (\ref{cp_potential_sphere}) for different truncation values of the angular momentum $l{\mbox{\tiny max}}=10,20,50,100,500,800$. This shows that truncation in $l$ is an especially delicate problem if the atom is in close proximity to the sphere, where the error induced by truncation of the $l$-series is largest. The green line is the full solution (with $l_{\mbox{\tiny max}}=800$) up to where the convergence to the half space is $U_{\mbox{\tiny BNR}}(r_l)/U(r_l)\simeq 3\%$, there it is replaced with the half-space potential.}
\label{fig:cp_fin}
\end{figure}

Given the material response functions \eqref{permit} and the tabulated optical data in Tables \ref{TAB:fit} and \ref{data_D}, we now have everything needed to calculate the CP potential Eq.~(\ref{cp_potential_sphere}) of an indium atom near a silicon dioxide sphere. However, the sum over spherical harmonics cannot in general be done analytically, so the series has to be truncated at value of $l$ large enough to keep errors within acceptable bounds.  Calculating the potential for extremely large $l$ is very time consuming computationally, so, based on the desired accuracy of our simulations, we decide upon a point at which the potential can be replaced by its half-space asymptote Eq.~\eqref{bnr}. We choose this accuracy to be at the $~3\%$ level, as beyond this the errors in the material response functions would dominate.  Carrying out this replacement procedure one finds the CP potential shown in Fig.~\ref{fig:cp_fin}. 

\subsection{CP-induced phase-shift in matter-wave diffraction} \label{MWScatteringPoissonSpotSection}
In this section we discuss the impact of the CP interaction on matter-wave diffraction, in particular on the Poisson spot. The Poisson spot is a bright spot which appears in the shadow region of a circular or spherical object due to diffraction. An approximate analogy with the double-slit experiment can be made by realizing that a circular (or spherical) diffracting object may be thought of as pairs of double slits arranged around a circle. The central maximum of the diffraction pattern for each pair of slits is on the axis, resulting in a bright spot. In other words, its appearance can be understood by the fact that the atomic paths from the point-source via the rim of the spherical object to any specific point on the optical axis all have the same length. The quantum-mechanical phases of the atoms thus positively interfere at the optical axis which results in Poisson's spot.

In order to quantify the effects of the CP potential on the Poisson spot we will use the Wentzel--Kramers--Brillouin (WKB) approximation, where the potential is assumed to change slowly relative to the de Broglie wavelength associated with the matter wave. Explicitly, the WKB approximation holds if the spatial derivative of the position-dependent wave vector $k(x)$ satisfies
\begin{equation}
k'(x) \ll k^2(x)
\end{equation}
which can be recast as
\begin{equation} \label{WKBCondition}
\frac{d}{dx} \sqrt{2m (E-U(x))} \ll \frac{2m(E-U(x))}{\hbar}
\end{equation}
where $E=\frac{1}{2} mv^2$ is the kinetic energy a particle of mass $m$ and velocity $v$, and $U$ is the potential it is subject to. To check that the approximation is valid here we consider an indium atom ($m=114.8$u) and SiO$_2$, as discussed in the previous section. For these materials we have the large-sphere non-retarded potential $U = U_{\mbox{\tiny BNR}}(r) = {C_3}/{z^3}$ given by Eq.~\eqref{bnr} with 
\begin{equation}\label{C3EstimateforWKB}
C_3 =\frac{\hbar }{16\pi^2\varepsilon_0} \inte{0}{\infty}{\xi}\alpha(\mathrm{i}\xi)\frac{\varepsilon(\mathrm{i}\xi)-1}{\varepsilon(\mathrm{i}\xi)+1} \approx 9.77 \times 10^{-50} \text{Jm}^3. 
\end{equation}
Using this potential and an approximate velocity of the indium atoms of $500$ m/s (see section \ref{ExperimentSection}) in Eq.~\eqref{WKBCondition}, one finds that for nanometer distances the left-hand-side is approximately nine orders of magnitude smaller than the right-hand side, meaning that for (at least) the case of the large-sphere non-retarded potential we are comfortably within the conditions of validity of the WKB approximation.

\begin{figure}
\centering
\includegraphics[width=\linewidth]{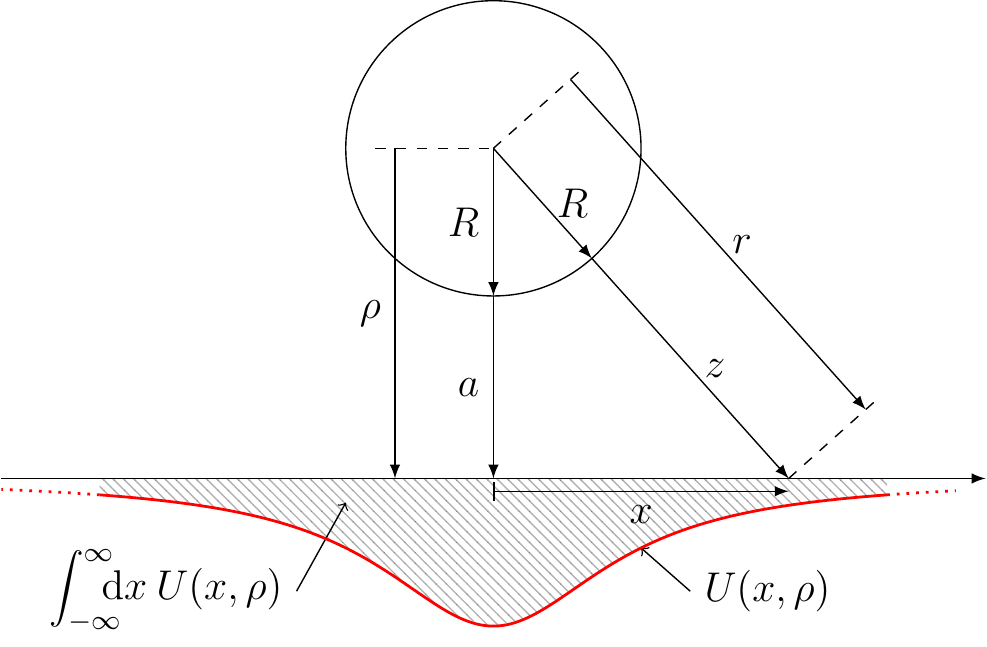}
\caption{Sketch of the lateral CP potential, this curve corresponds to the CP interaction of an atom flying in a straight line parallel to the optical axis nearby the sphere surface. }
\label{fig:lat_cp}
\end{figure}

We describe the trajectory in our specific system through the co-ordinates $x$ and $\rho=a+R$ as indicated in Fig.~\ref{SphereDiagram}. This means that the CP-induced phase shift  $\Delta \varphi_{\mbox{\tiny CP}}$ (also known as the eikonal phase) is given by \cite{Perreault2005}:
\begin{equation}
\Delta\varphi_{\mbox{\tiny CP}}(\rho)=- \frac{1}{\hbar v}\inte{-\infty}{\infty}{x}U(x,\rho)
\label{phase_shift}
\end{equation}
Given this phase shift, one can then calculate a diffraction pattern using the Fresnel approximation (where the wavelength of the beam undergoing diffraction is much smaller than the dimensions of the diffracting object). The amplitude $A(P)$ of the signal at a point $P$ in the image plane (see Fig. \ref{fig:fresnelsimulation}) is given by \cite{Dauger1996,reisinger_particlewave_2011}:
\begin{equation}
A(P)=-\frac{i}{\lambda gb}\inte{0}{2\pi\!\!\!}{\phi}\!\!\!\inte{0}{\infty}{\rho} G(\phi,\rho) \rho e^{i[\varphi_g(\rho)+\Delta\varphi_{\mbox{\tiny CP}}(\rho)]}\:
\label{integr_fresnel}
\end{equation}
where $\varphi_{g}(\rho)=\frac{\pi}{\lambda} (\frac{1}{g} + \frac{1}{b})\rho^2$ is the phase-shift induced by the geometry of the object in Fresnel approximation and $\varphi_{\mbox{\tiny CP}}(\rho)$ is the CP-induced phase shift given by Eq.~\eqref{phase_shift}
with $U(x,\rho)$ the lateral CP potential (see Fig.~\ref{fig:lat_cp}). The function $G(\phi,\rho)$ is the aperture function representing the circular cross-section of the sphere. It is $0$ for points that are located within the blocked cross-section and $1$ otherwise. To calculate the amplitude for an arbitrary point $P$ in the detection plane the origin used in the integral \eqref{integr_fresnel} is shifted to the intersection point of the line connecting the source and image points with the integration plane, resulting in the new radial coordinate $\overline{\rho}$. The numerical evaluation of $|A|^2$, which is equal to the intensities in the imaging plane, is discussed in section \ref{ExperimentSection}.

\begin{figure}
\centering
\includegraphics[width=0.9\columnwidth]{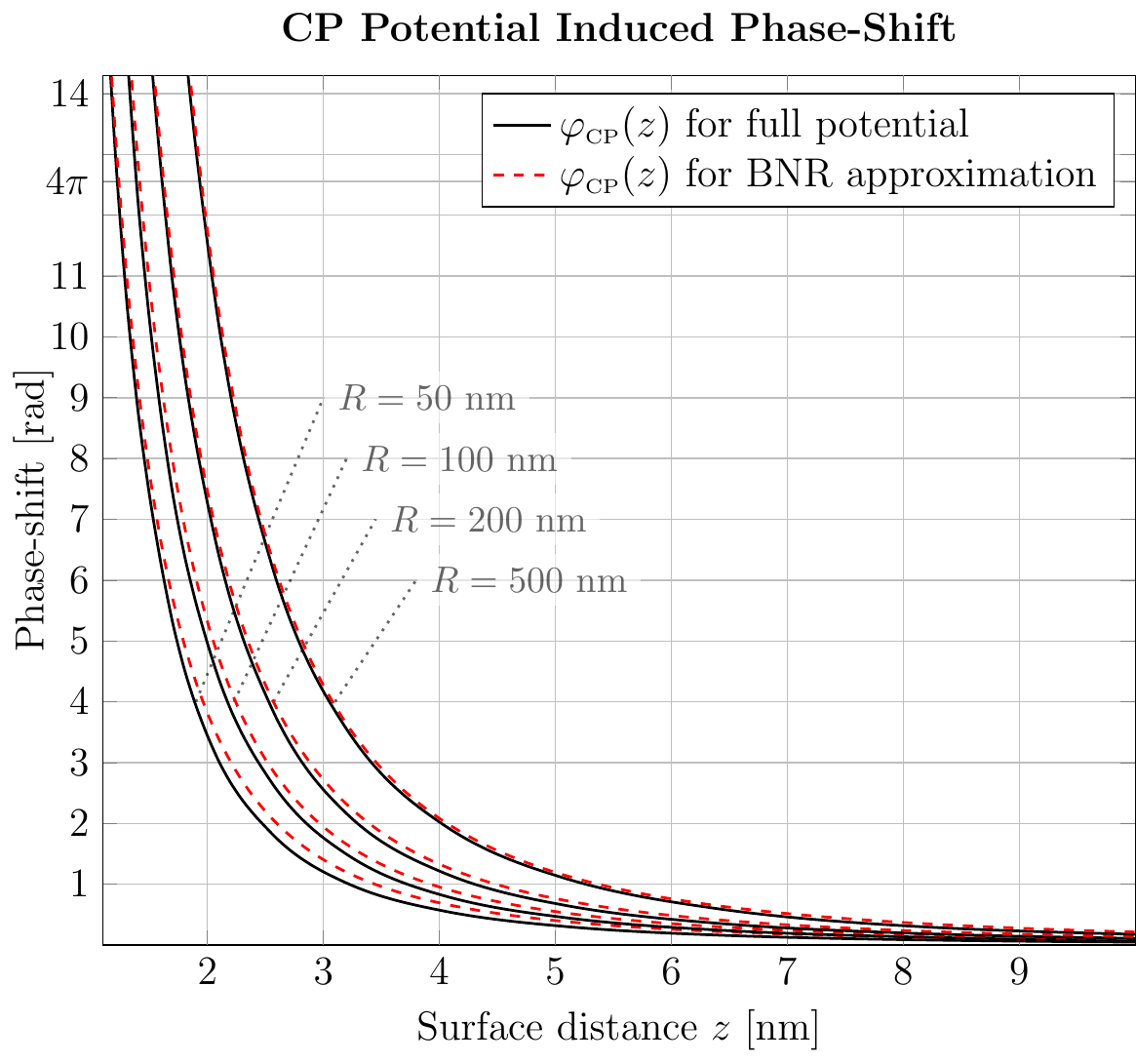}
\caption{(Color online). CP potential-induced phase shift for spheres with different radii ($50$, $100$, $200$ and $500$ nm). The black line is the phase-shift where the full solution for the CP potential from Eq.~\eqref{phase_shift} has been used. The red dashed line is the phase-shift for the leading order of the half-space approximation. Note that the approximation improves for larger spheres, in line with the intuition that a sphere with a larger radius is more similar to a half-space.}
\label{fig:cp_phase_shift}
\end{figure}
The Fresnel approximation is accurate in the discussed experiment since the object and image distances $g$ and $b$ are large compared to the size of the diffraction object $R$ and the wavelength $\lambda$ is much smaller than $R$. In addition, note that although Fresnel theory only applies to two-dimensional objects, the ``volume'' of the sphere is implicitly taken into account by the accumulated phase-shift Eq.~(\ref{phase_shift}). 

For distances of less than approximately $10$ nm 
the phase shift is well-approximated by the large-sphere potential of Eq.~\eqref{bnr}, as shown in  Fig.~\ref{fig:cp_phase_shift}. The potential takes a particular simple form in this limit \cite{Hornberger2012}, namely:
\begin{equation}
U_{\mbox{\tiny BNR}}(r) = -\frac{C_3}{z^3}=-\frac{C_3}{\left[\sqrt{(a+R)^2+x^2}-R\right]^3}
\end{equation}
as shown in Fig.~\ref{fig:lat_cp}.  This means that if this potential is used in the phase shift integral, the result can in fact be found analytically. 
\begin{align}
&\Delta \varphi_{\mbox{\tiny CP, BNR}}(a)=\frac{C_3 }{2 \hbar v}\frac{1}{{a^2 (2 R+a)^2}}\Bigg\{{6 R^2+8 R a+4 a^2}\notag \\
&+\frac{3 R (R+a)^2}{\sqrt{a(2R+a)}} \left[2 \arctan\left(\frac{R}{\sqrt{a (2 R+a)}}\right)+\pi \right]\Bigg\}\notag \\
&\approx \frac{ C_3}{2\hbar v} \frac{3 \pi \sqrt{R}}{2\sqrt{2} a^{5/2}} \quad \text{for } a\ll R \label{eq:varphi}
\end{align}
with $C_3\approx 9.77 \times 10^{-50} \text{Jm}^3$ as explained above. Finally we define for later convenience the quantity $  C_{52}  \equiv \frac{ C_3}{2\hbar v} \frac{3 \pi \sqrt{R}}{2\sqrt{2} }$ so that
\begin{equation}\label{C52Def}
\Delta \varphi_{\mbox{\tiny CP, BNR}}(a \ll R) = \frac{C_{52}}{a^{5/2}} \, .
\end{equation}

\section{ Calculation of diffraction images}
\label{ExperimentSection}

In this section the derived Casimir-Polder phase shift is used in a numerical solution of the Fresnel diffraction integral in order to predict the effect of the Casimir-Polder interaction on the relative intensity of Poisson's spot. With relative intensity $I_{rel}$ we refer to the ratio between the intensity at the center of Poisson's spot in the detection plane and the intensity of the undisturbed beam, also in the detection plane. These predicted relative intensities can then be compared to intensity data from the aforementioned matter-wave experiments. 

\begin{figure}
\centering
\includegraphics[]{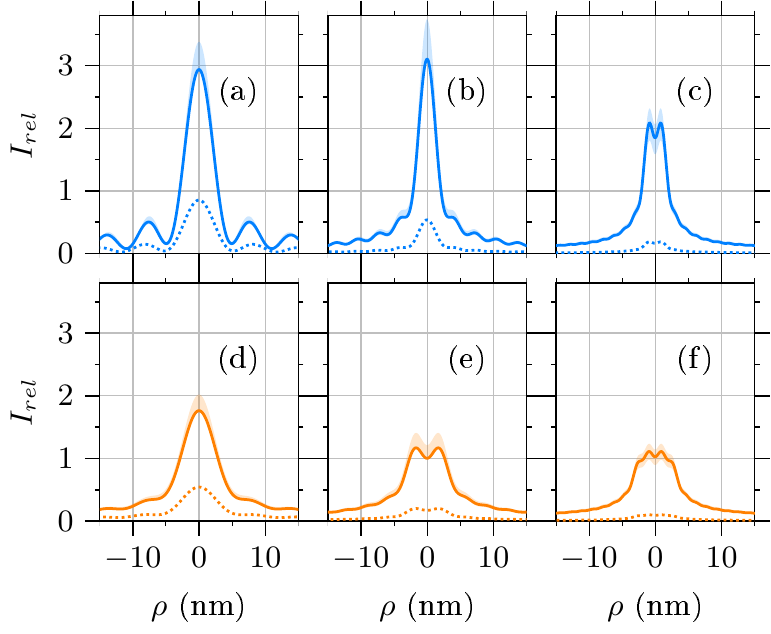}
\caption{(Color online.) Lateral relative intensity at $b=0.1$ mm assuming a source diameter of $20~\mu$m  (a-c) and $40~\mu$m (d-f). The assumed sphere diameter is  $R=50$ nm (a,d), $R=100$ nm (b,e), $R=200$ nm (c,f). The model including the large-sphere non-retarded CP-phase-shift is depicted using continuous lines and the Fresnel-diffraction-only model using dotted lines. The shaded regions around the solid lines give an approximate error margin stemming from the formation of a thin indium layer on the sphere (see section \ref{ThinFilmSection}).}
\label{fig:lateral}
\end{figure}

The parameters assumed in the calculations, and detailed in the following lines, are chosen according to the setup used in the experiment. The oven source consists of a closed molybdenum crucible with a nominally $20\mu$m-diameter orifice and is kept at a temperature of $T_s=1200\degree$C. The temperature is chosen this high to generate a substantial partial pressure of indium (about $87$ Pa) within the crucible, resulting in the high source-brightness needed to observe Poisson's spot. The orifice diameter is also small enough to avoid any increase in the virtual source size \cite{reisinger2007,Reisinger2012}. In spite of the relatively high pressure, the speed of the exiting indium atoms is expected to be characterized by the thermal speed distribution inside the source, with an approximate mean velocity given by $\bar{v} = \sqrt{8 k_b T_s/(\pi m)} \approx 521$ m/s, where $k_b$ is Boltzman's constant and $m$ is the atomic mass of indium ($m=114.8$ u). This corresponds to a mean de-Broglie wavelength of $6.67$ pm, which is the wavelength used in the calculations described below. The speed of the atoms affects the relative intensity in two ways: (1) Higher atomic speed results in smaller wavelengths and thus in a thinner point-source Poisson spot which is equivalent to lower relative intensity for extended sources. (2) Higher atomic speed also results in shorter Casimir-Polder interaction times and thus a reduced phase shift, which also results in a reduction of the relative Poisson spot intensity \cite{Juffmann2012b,reisinger_particlewave_2011}, as can be seen below. The spread in wavelengths is neglected as its effect on the relative intensity of Poisson's spot is expected to average out. A clear sign of the wave nature are the side maxima (as for example visible in Fig.~\ref{fig:lateral}), unlike the Poisson spot itself which has a classical analogue due to particle deflection in the CP potential \cite{reisinger_particlewave_2011,Juffmann2012b}. The visibility of these side maxima is, however, affected by the spread in wavelengths, which are therefore hard to detect in practice.  Three different sphere diameters of the silicon-dioxide particles will be assumed ($R= 50$ nm, $100$ nm, $200$ nm) and a fixed distance between the source and the sphere of $g=600$ mm. The image distance $b$ between the sphere and the detection plane is varied in the range $b=0.05-1.05$ mm. 

The disturbance $A(P)$ at a point $P$ in the detection plane can be expressed by Eq.~\ref{integr_fresnel}, which makes use of the Fresnel approximation and already incorporates the phase shift expected from the Casimir-Polder interaction \cite{Cronin2005}. 

\begin{figure}
\centering
\includegraphics[]{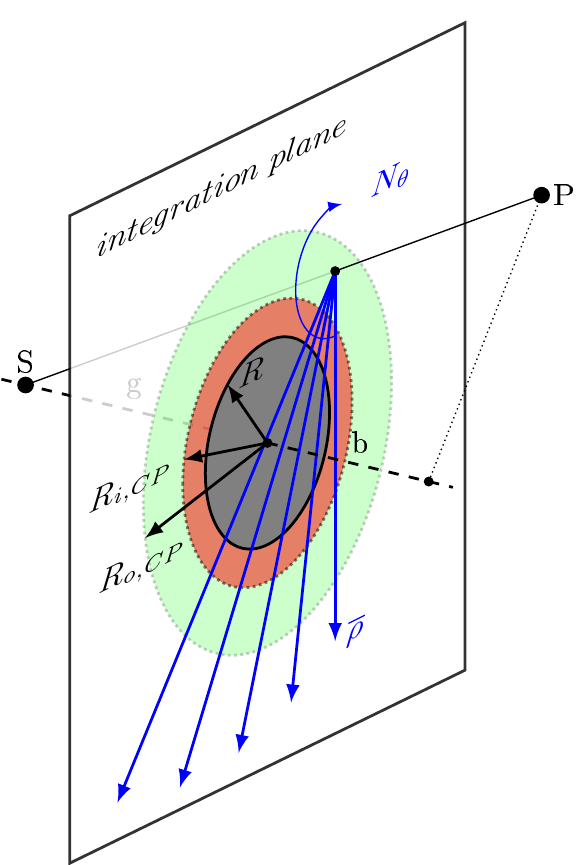}
\caption{(Color online.) Schematic indicating numerical solution of Fresnel integral. In order to calculate the amplitude $A(P)$ at point $P$ due to a point source at $S$ the surface integral from equation (\ref{integr_fresnel}) in the indicated integration plane needs to be solved. Note that to calculate the amplitude for an arbitrary point $P$ in the detection plane the origin used in the integral is shifted to the intersection point of the line $\overline{SP}$ and the integration plane, resulting in the new radial coordinate $\overline{\rho}$. The circular cross-section of the sphere is shown in gray with a solid circumference. The adjacent annular zone in orange with an outer radius $R_{i,\text{CP}}$ indicates the region where the CP/vdW phase shift is larger than $4\pi$. This zone was therefore neglected entirely in any of the radial line integrals, which are indicated by the blue arrows. Further away from the optical axis the annular zone shown in green corresponds to the region where the CP/vdW phase shift is in the range [$\frac{\pi}{1000}$, $4\pi$] with an outer radius of $R_{o,\text{CP}}$. In this region each radial line integral was solved numerically, taking into account the surface distance $a$ at each numerical integration step. Outside of it, the CP/vdW phase shift is neglected and the contribution to the line integrals simplifies to two terms corresponding to the  intersection point(s) with the circle of radius $R_{o,\text{CP}}$ with the radial integration line and a point at infinity (see Ref. \cite{Dauger1996}).
}
\label{fig:fresnelsimulation}
\end{figure}

The phase-shift $\varphi_{CP}(r)$ is only non-negligible in an annular region in the integration plane between radii $R_{i,\text{CP}}$ and $R_{o,\text{CP}}$ (see Fig.~\ref{fig:fresnelsimulation}). Very close to the sphere the phase shift starts to oscillate increasingly fast as a function of $z$. It is safe to neglect contributions originating from an annular region of radius $R_{i,CP}$ and inward - i.e. immediately adjacent to the sphere. This is because, from a classical point of view, trajectories passing within $R_{i,CP}$ result in large particle deflections or even particle capture by the sphere and thus do not contribute to the diffraction image close to the optical axis. In the calculations presented here $R_{o,\text{CP}}$ and $R_{i,\text{CP}}$ are set such that the phase shift equals $\pi/1000$ and $4\pi$, respectively (this turns out to be more efficient and accurate than the absolutely fixed boundaries used in ref. \cite{reisinger_particlewave_2011}). For the sphere radii $R=50$ nm, $100$ nm, and $200$ nm, for which results are reported below, the phase shift constants defined by Eq.~\eqref{C52Def} are $C_{52}=6.622\cdot 10^{-22}$ m$^{5/2}$, $9.365\cdot 10^{-22}$ m$^{5/2}$, and $13.244\cdot 10^{-22}$ m$^{5/2}$, respectively, and the boundary radii are  $[R_{i,\text{CP}},R_{o,\text{CP}}]=[51.2,83.8]$ nm,  $[101.4,138.9]$ nm, and $[201.6,244.7]$ nm, respectively, for the given beam parameters.

The surface integral is solved numerically following the general approach discussed in ref.
\cite{Dauger1996} and explained schematically in Fig.~\ref{fig:fresnelsimulation}. The integral is replaced by two sums. The first solves the integral in the $\theta$ variable, corresponding to $N_{\theta}$ radially equally spaced rays. We choose $N_{\theta} = 19\, 997$ (a prime number) to avoid artificial fringes from symmetry in the numerical evaluation.  The second sum, that corresponds to line integrals in the radial direction, reduces to a few summands that are evaluated at the intersection points of each particular ray with the edges of transmitting regions. In the annular region where the CP potential is non-negligible this simplification does not hold. Therefore, whenever, a ray traverses this region the corresponding part of the radial line integral is  computed using a simple trapezoidal rule, taking into account the local phase shift. The resolution of this numerical line integration was fixed at $0.1$ nm. For each image distance $b$ and sphere radius $R$ the intensities corresponding to a row of $2000$ pixels reaching from the optical axis to the radius $R$ in the image plane is computed with this method. The complete $4000\times4000$ pixel 2d point-source diffraction image is inferred from symmetry and interpolation. Finally, the image is convoluted with the demagnified image of the source, of width $20\cdot b/g~\mu$m$\cdot$.

\begin{figure}
\centering
\includegraphics[]{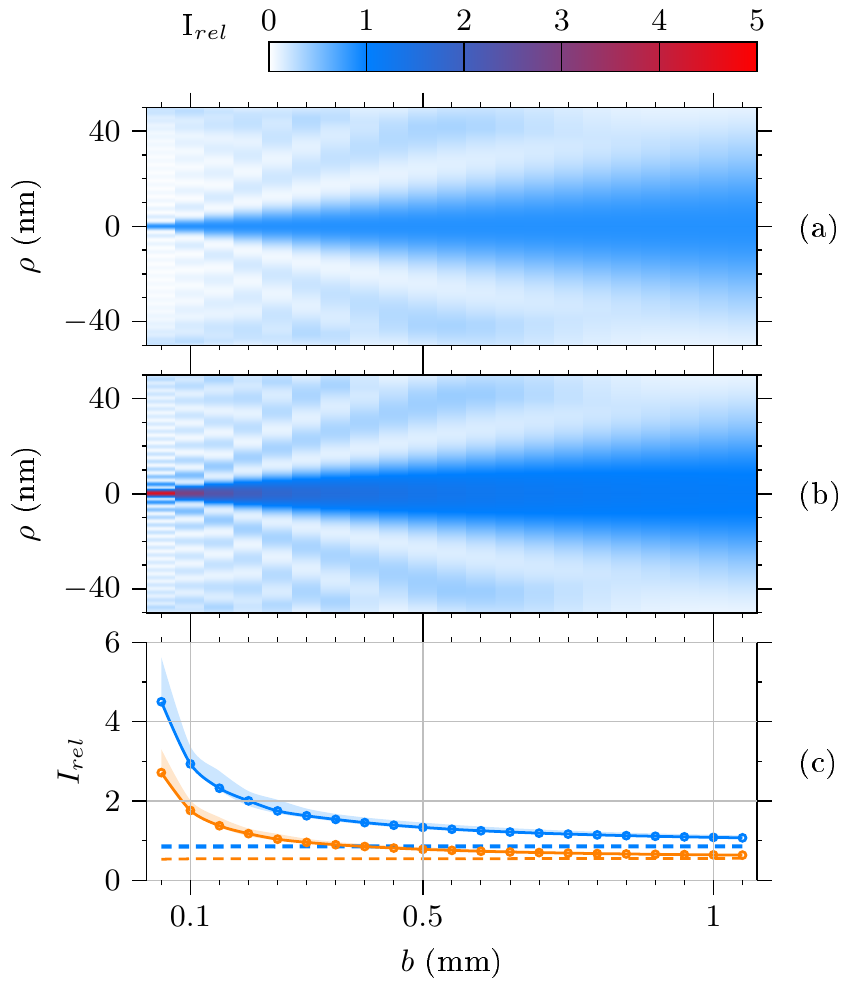}
\caption{(Color online.) Relative diffraction intensity for sphere radius $R=50$ nm. The plots in this figure show the calculated relative intensity of an atomic indium beam in the shadow region behind a silicon-dioxide sphere with a diameter of $100$ nm. The formation of Poisson's spot about the optical axis (horizontal) can be clearly seen. The beam originates from a $20-\mu$m-diameter source at a distance of $600$ mm from the sphere. $I_{rel}$ is given in intensity units of the undisturbed wave front (Without the sphere the plot would show $I_{rel}=1$). Here $\rho$ denotes the distance from the optical axis and $b$ is the distance between sphere and detection plane. For comparison we display results for the case of no interaction between the sphere and the beam (a) and including the Casimir-Polder induced phase shifts using the large-sphere non-retarded potential (b). In (c) the relative intensity on the optical axis is shown. The dashed and continuous lines shown in blue correspond to the data in (a) and (b), respectively, which means that the solid lines include the CP potential, while the dashed lines do not. The lines in orange show the trend for the same parameters, but assume a $40-\mu$m-diameter beam source. The shaded region about the solid lines give an approximate error margin due to a thin layer of indium forming on the sphere (see text).  }
\label{fig:sphere100}
\end{figure}

\begin{figure}
\centering
\includegraphics[]{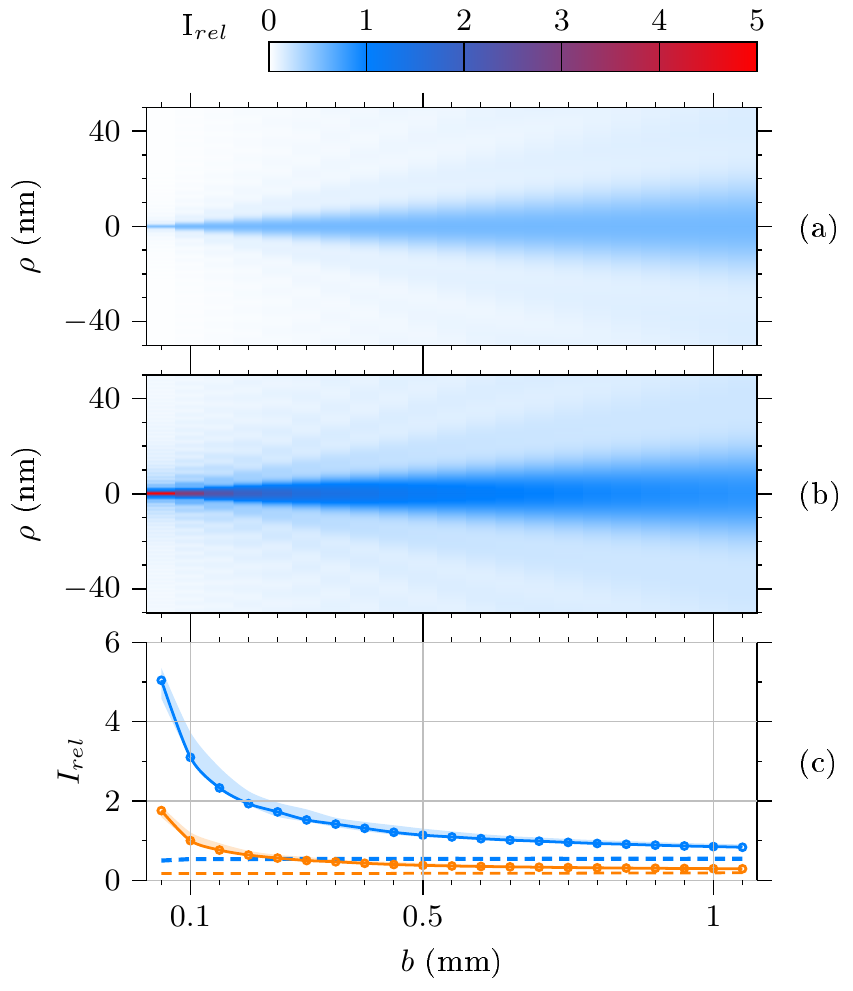}
\caption{(Color online.) Relative diffraction intensity for sphere radius $R=100$ nm. The graphs are analogous to figure \ref{fig:sphere100}.}
\label{fig:sphere200}
\end{figure}

\begin{figure}
\centering
\includegraphics[]{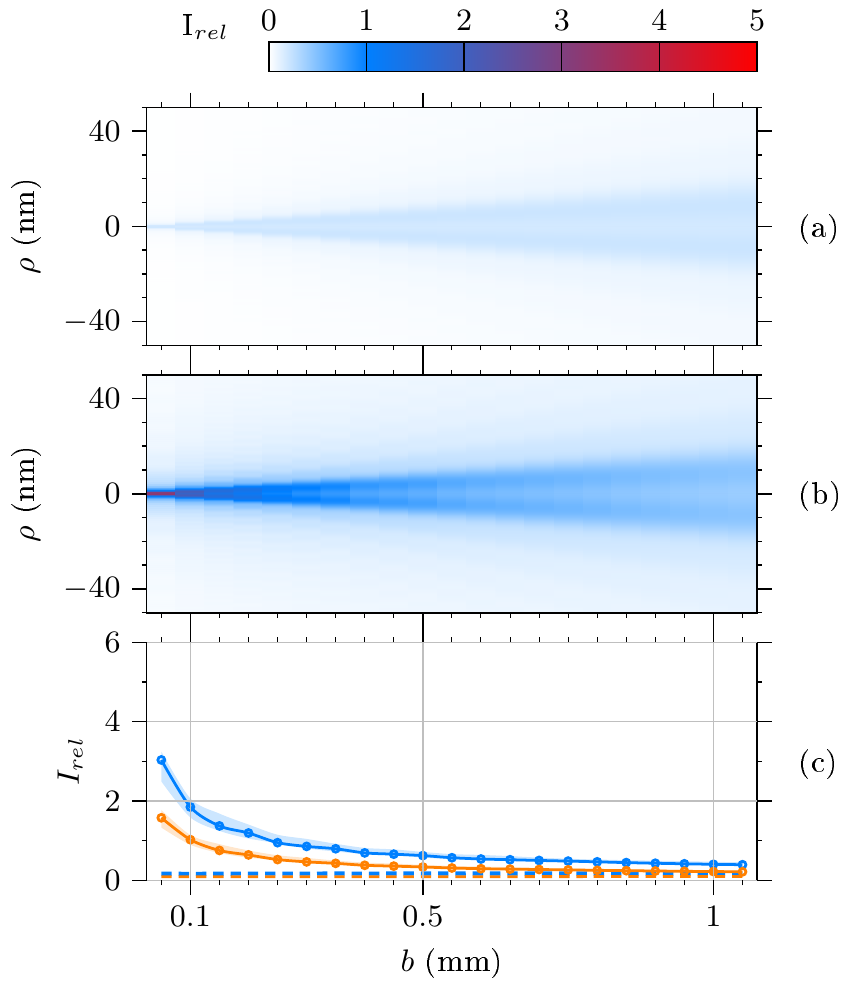}
\caption{(Color online.) Relative diffraction intensity for sphere radius $R=200$ nm. The graphs are analogous to figure \ref{fig:sphere100}.}
\label{fig:sphere400}
\end{figure}

\section{Discussion and Outlook}
\label{DiscussionSection}

As can be seen in Fig.~\ref{fig:cp_phase_shift} the change in the phase shift due to retardation or the size of the sphere is negligible for the experiment discussed here. For this reason we have limited the Fresnel diffraction simulations to the simpler half-space, non-retarded approximation.

The resulting relative intensities as a function of $\rho$ and $b$ are plotted in Figs \ref{fig:sphere100}, \ref{fig:sphere200} and \ref{fig:sphere400} for three different sphere diameters. For better comparison the lateral relative intensity distributions are shown at the image distance $b=0.1$ mm in Fig. \ref{fig:lateral}. The relative intensity of Poisson's spot is increasingly amplified at smaller distances $b$ due to the CP interaction. In addition a small shift of the side maxima toward the optical axis can be noted (see especially Fig. \ref{fig:lateral}(a)), which we attribute to an increasing effective sphere diameter for stronger CP interaction. The plot of the on-axis intensity for two different source sizes shows that increased spatial coherence leads to a more pronounced sensitivity of the Poisson spot intensity to the CP potential. By comparing Figs \ref{fig:sphere100},\ref{fig:sphere200}, and \ref{fig:sphere400} one can see that an increase in sphere diameter both increases $I_{rel}$ due to the longer time the particle spends in the vicinity of the sphere, but also decreases it as expected from Fresnel diffraction. In other words there are two competing effects, which is why $I_{rel}$ is at a maximum for the medium sphere diameter (only in case of the $20-\mu$m source). 

To ensure the reliability of our results, we compared them to those found using a completely different numerical approach. As discussed in the caption of Fig. \ref{fig:fresnelsimulation}, the results plotted in Figs. \ref{fig:sphere100}, \ref{fig:sphere200}, \ref{fig:sphere400} and \ref{fig:lateral} were computed in a similar way to \cite{Dauger1996}, i.e. by direct numerical implementation of the Fresnel integral. That method is equivalent to the phase-space treatment outlined in \cite{Juffmann2012b} using Wigner functions (see \cite{Nimmrichter2014} for details). The phase-space framework is ideally suited to account for environmental decoherence effects \cite{Hornberger2003,Hornberger2004a}, e.g. by background gas collisions, and to juxtapose the predictions of the matter-wave model and a classical ballistic treatment of the atom trajectories. We have checked our numerical results against this framework and find agreement at the percent level, with the dominant contribution to the difference being our use of the approximate expression in the final line of \eqref{eq:varphi}. Another consistency check between our work and that of \cite{Juffmann2012b} is that the latter can predict which (semi)-classical trajectories physically collide with the sphere due to deflection by the potential. This can be determined simply by imposing conservation of energy and momentum, then minimising the resulting function to find the smallest impact parameter $a_\text{min}$ that escapes the potential. For the cases $R=50\,$nm, $100\,$nm, $200\,$nm considered here, we find $a_\text{min}=1.0\,$nm, $1.2\,$nm, and $1.4\,$nm, respectively, which is consistent with the values $R_{i,\text{CP}}$ \footnote{Note that the $R_{i,\text{CP}}$ values quoted in sec.~\ref{ExperimentSection} include the sphere radii R.} derived from our phase criteria in section \ref{ExperimentSection}.

There are three more effects that we have not addressed so far, but which we discuss in the following subsections. 

\subsection{Surface Corrugation}
The calculation neglects any surface corrugation of the sphere, for which a reduction in Poisson spot intensity is expected at small distances $b$ behind the sphere from the zero-interaction Fresnel-Kirchhoff integral. This effect can be estimated using an analytic dampening factor \cite{reisinger_relative_2016} that can be applied to the on-axis intensities. The relative intensity of Poisson's spot will be close to zero if the amplitude of the surface corrugation is approximately equal to the width of the adjacent Fresnel zone $w_{fz} = \sqrt{R^2 + \frac{\lambda~g~b}{(g+b)}} - R$. Assuming a corrugation amplitude of about $1$ nm, we have $w_{fz} \approx 1$ nm at distances $b=0.015$, $0.03$ nm, and $0.06$ nm for sphere radii R=$50$ nm, $100$ nm, and $200$ nm, respectively. A corrugation amplitude of $10$ nm entails approximately a 10-fold increase in the values of $b$ at which the Poisson spot is no longer visible. This illustrates the importance of avoiding surface corrugation in the experiment as much as possible.

Furthermore, surface corrugation can influence the CP potential in the vicinity of the sphere in non-trivial ways \cite{Henkel1998,Messina2009,Contreras-Reyes2010,Bender2014a,Bennett2015,Buhmann2016}. In practice, we expect that the presence of CP interaction effectively mitigates the requirements on surface corrugation to some degree, especially if the corrugation amplitude is less than $R_{i,\text{CP}}-R$. Accurate accounting of this influence could help in the future to distinguish between quantum and classical behavior of mesoscopic particles \cite{reisinger_particlewave_2011}. The details of this, however, we anticipate to be an interesting route for further study.

\subsection{Formation of a metallic thin film on the sphere}\label{ThinFilmSection}
\begin{figure}
\centering
\includegraphics[width=\linewidth]{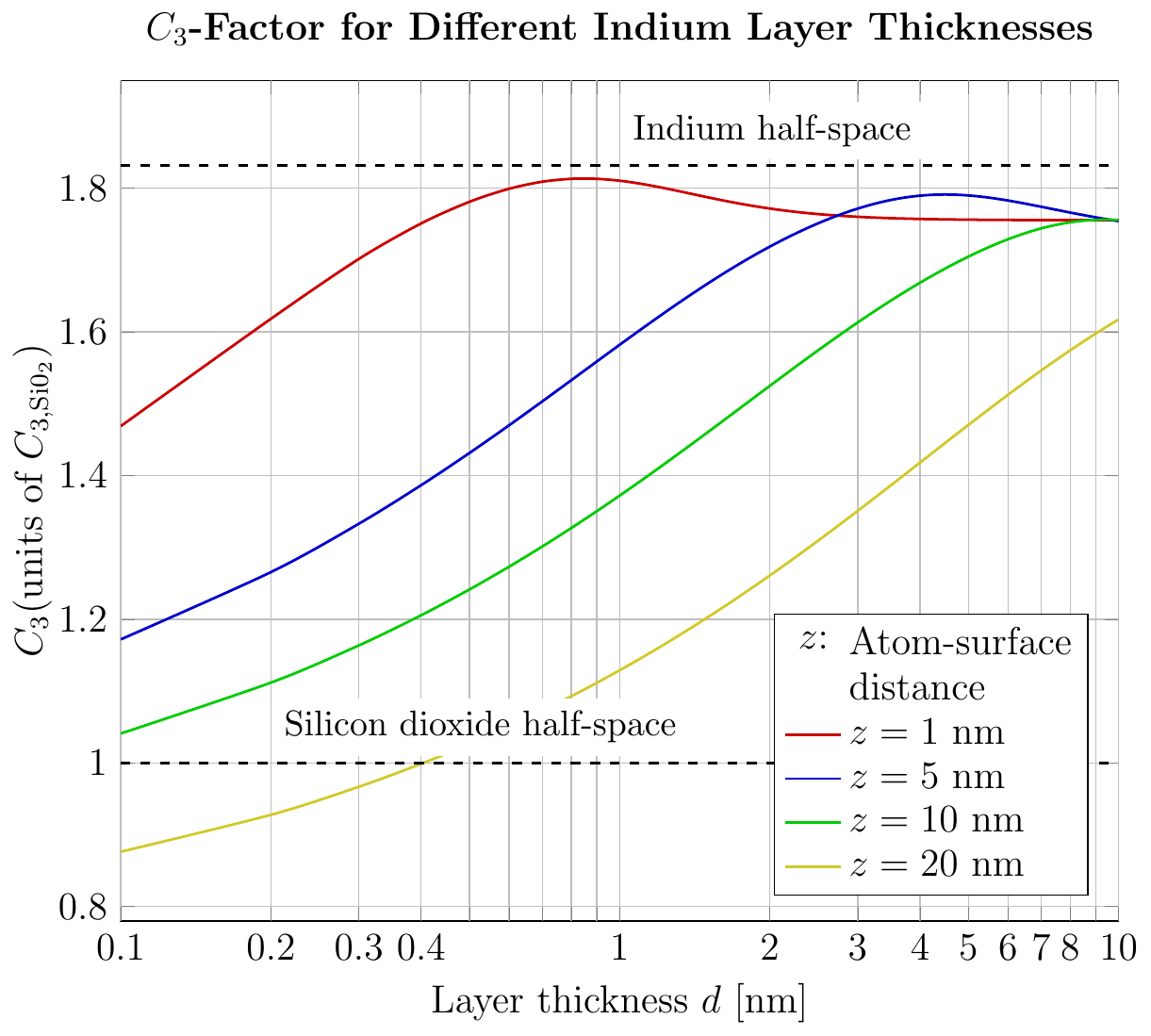}
\caption{(Color online.) Variation of the non-retarded, half-space CP constant $C_3$ as a function of deposited indium film thickness. }
\label{fig:layeredindium}
\end{figure}

One more reason for a deviation of experimental data from the results presented above is the possible buildup of an indium film on the silicon dioxide sphere. In our large-sphere approximation this would manifest itself as a thin layer deposited on top of a half-space, for which estimates of its influence on the effective Casimir-Polder potential can be obtained relatively easily. We present a preliminary investigation of this in Fig.~\ref{fig:layeredindium}, where an effective $C_3$ at various distances from the coated sphere is shown as a function of indium film thickness. For thin layers a screening effect can be noted far from the surface. As the film grows in thickness the half-space CP potential of a pure indium surface is reached. The graph suggests that the effect can be accounted for by a modification of  the effective $C_3$ of the system by approximately a factor in the range $0.8$--$1.8$ (depending on layer thickness). This would result in an exposure-time dependent diffraction pattern as the indium continuously accumulates upon the sphere. We show the deviation in the relative intensity of Poisson's spot approximately possible due to the thin film in the form of an error corridor (see shaded region in the plots of figures \ref{fig:sphere100}-\ref{fig:lateral}). The boundaries of the corridors were calculated by assuming a constant effective CP constants of $0.8\,C_3$ and $1.8\,C_3$. The variation of the effective $C_3$ as a function of distance from the sphere  $z$, as predicted in Fig.~\ref{fig:layeredindium}, can lead to even stronger attenuation or amplification depending on the resulting CP phase shift relative to the geometrical phase shift. However, we expect the error to be of the same order of magnitude as depicted by the shaded error corridors. The results suggest that the change in intensity of Poisson's spot due to a metallic thin film can be observed, but maybe in practice hard to quantify. The main reasons for this are additional modifications of $I_{rel}$ to be expected from changing surface corrugation as the thin film is deposited. 

\subsection{Temperature}

\subsubsection{Temperature of indium atoms}

The effect of the CP interaction on the relative intensity of Poisson's spot depends on the temperature $T_s$ of the indium atoms, at which they emerge from the oven source, in two distinct ways. First, $T_s$ determines the speed distribution of the atoms and thus the accumulated CP phase shift (see equation \ref{eq:varphi}). The speed distribution also affects the geometrical Fresnel phase shift via the de-Broglie wavelength of the atoms. Both of these manifestations of finite $T_s$ have been accounted for in the presented calculations. Second, at $T_s$ any number of internal degrees of freedom of the atom maybe excited, which would alter the atom's polarizability and thus the CP potential. The occupation probability of the lowest excited state of the indium atoms at temperature $T_s=1200\degree$C can be estimated using the Boltzmann factor $e^{-\frac{\hbar \omega_{01}}{k_B T_s}}  $ with the transition frequency $\omega_{01}$ from table \ref{data_D} and the reduced Planck constant $\hbar$ and Boltzman's constant $k_B$. This evaluates to approximately $5\cdot 10^{-11}$, which makes the assumption that all indium atoms reside in the ground state an extremely good one. Even at higher temperatures accessible with a standard oven heater, this ratio remains negligible. However, an artificial excitement using a laser at one of the specific indium wavelengths could be an appealing route to probing the CP interaction with excited atoms.

\subsubsection{Ambient temperature}

The ambient temperature of the experimental apparatus floods the interaction region between atom and dielectric sphere with thermal photons. These additional excitations of the electromagnetic field affect the CP potential only at distances of the order of the wavelengths of the thermal photons \cite{Gorza2006,ChavesdeSouzaSegundo2007,Henkel2002} (approximately 48 $\mu$m at room temperature). Since we determined that the CP phase shift in the discussed experiment is completely negligible at distances exceeding about $50$ nm, it is safe to ignore any contributions from ambient thermal photons. 

\subsubsection{Temperature of the silicon-dioxide sphere}

While it is not practicable to change the temperature of the apparatus significantly, the temperature of the diffraction obstacle could be raised to about $1000\degree$C, and higher for alternative obstacle materials. A reason for heating the obstacle could be to prevent the deposition of a thin film of the beam species, as discussed above. This would result in immediate re-evaporation of beam particles captured by the sphere, reflecting them diffusely in the general direction of the source. The influence of such states of thermal non-equilibrium on the CP potential is a topic of current research \cite{Antezza2004,Obrecht2007,Buhmann2008} and its possible influence on the present experiment should be the subject of further study.

\section{Conclusion}
\label{ConclusionSection}

We have presented a detailed treatment of the CP potential between indium atoms and a silicon-dioxide sphere and its influence in the case of Poisson spot matter-wave diffraction experiments. The main feature of our results is that the makeshift models of Casimir-Polder potentials, that neglect retardation and surface curvature, and were used so far in matter-wave diffraction experiments are in fact completely adequate. We have shown this by making a detailed account of the situation for a realistic and ongoing Poisson spot experiment. This has allowed us to make verifiable predictions of diffraction patterns and relative intensities of the Poisson spot, backed up by a proper account of geometry- and material-dependent dispersion forces. We found that the diameter of the silicon dioxide sphere mainly affects the relative intensity of Poisson's spot due to the related change in length of the interaction region. Furthermore, we have estimated the effect from surface corrugation of the silicon dioxide sphere and the possible deposition of indium on the diffraction obstacle. Finally, there remain a few more minor idealisations that are not included in our model thus far, for example that the sphere is at thermal equilibrium with the indium beam. On the whole we expect that the predictions for the relative intensity of Poisson's spot made here, provide solid ground for tests of the CP potential as predicted by macroscopic quantum electrodynamics in the ongoing experiments.

\section{Acknowledgements}
\label{AcknowledgementsSection}
We thank Stefan Scheel for fruitful discussions. S.Y.B and R.B. acknowledge support from the Deutsche Forschungsgemeinschaft (grant BU 1803/3-1), and S.Y.B. additionally acknowledges support from the Freiburg Institute for Advanced Studies (FRIAS). J.F. and S.Y.B. acknowledge support by the Research Innovation Fund of Freiburg University. T.R. acknowledges support by the Ministry of Science, Research and Art of Baden-W\"{u}rttemberg via a Research Seed Capital (RISC) grant. T.R., H.G. and H.H. acknowledge support by the Helmholtz Association. 

\appendix
\section{Vector wave functions}\label{Appendix}
The vector wave functions entering into \eqref{tens_sphere} are given by:
\begin{multline}
\mathbf{a}_{lm\pm TE}(\mathbf{r})= \mp h^{(1)}_l(k_jr)m\frac{P^m_l(\cos\theta)}{\sin\theta}\elec{\sin(m\phi)}{\cos(m\phi)}\mathbf{e}_\theta  \\
 + h^{(1)}_l(k_jr)\frac{d}{d\theta}\big(P^m_l(\cos\theta)\big)\elec{\cos(m\phi)}{\sin(m\phi)}\mathbf{e}_\phi,
\end{multline}
\begin{multline}
\mathbf{a}_{lm\pm TM}(\mathbf{r})= l(l+1)\frac{h^{(1)}_l(k_jr)}{k_jr}P^m_l(\cos\theta)\elec{\cos(m\phi)}{\sin(m\phi)}\mathbf{e}_r  \\
 + \frac{[k_jrh^{(1)}_l(k_jr)]'}{k_jr}\frac{d}{d\theta}\big(P^m_l(\cos\theta)\big)\elec{\cos(m\phi)}{\sin(m\phi)}\mathbf{e}_\theta  \\
 \mp \frac{[k_jrh^{(1)}_l(k_jr)]'}{k_jr}m\frac{P^m_l(\cos\theta)}{\sin\theta}\elec{\sin(m\phi)}{\cos(m\phi)}\mathbf{e}_\phi.
\end{multline}
and where the primes are to be understood in the same sense as detailed below Eq.~\eqref{mie_refl}. The notation $\scriptsize{\elec{x}{y}}$ here means that its upper or lower entries should be taken consistent with corresponding entries of the relevant $\pm$. 

\section{Green's tensor simplifications}\label{GTAppendix}

The trace of the scattering Greens's tensor \eqref{tens_sphere} reads:
\begin{align}
\label{gen_tens_sph}
&\Tr{\tens{G}^{(1)}(\mathbf{r},\mathbf{r},i\xi)}\!=\! -\frac{\xi}{4\pi c}\sum\limits_{l=1}^{\infty}\!\sum\limits_{m=0}^{l}(2-\delta_{m0})\frac{2l+1}{l(l+1)}\frac{(l-m)!}{(l+m)!}\notag \\
&\times \Bigg\{r_{lTE}[h^{(1)}_l(kr)]^2\bigg[m^2\bigg[\frac{P^m_l(\cos\theta)}{\sin\theta}\bigg]^2+\bigg[\frac{dP^m_l(\cos\theta)}{d\theta}\bigg]^2\bigg]\notag\\
&+r_{lTM}\bigg[\bigg(l(l+1)P^m_l(\cos\theta)\frac{h^{(1)}_l(kr)}{kr}\bigg)^2 + \frac{[krh^{(1)}_l(kr)]'^2}{(kr)^2}\notag\\
&\times\bigg(m^2\bigg[\frac{P^m_l(\cos\theta)}{\sin\theta}\bigg]^2+\bigg[\frac{dP^m_l(\cos\theta)}{d\theta}\bigg]^2\bigg)\bigg]\Bigg\}.
\end{align}
To carry out the sum over $m$, we use the addition theorem for spherical harmonics \cite{jackson1998classical}:
\begin{multline}
\label{spher_sum_theo}
P_l(\cos\gamma)=
\sum_{m=0}^{l}(2-\delta_{m0})\frac{(l-m)!}{(l+m)!}\\\times P_l^m(\cos\theta)P_l^m(\cos\theta')\cos(m(\phi-\phi')).
\end{multline}
The trace can therefore be rewritten as:
\begin{multline}
\label{eq:simp_trace_G}
\Tr{\tens{G}^{(1)}(\mathbf{r},\mathbf{r},i\xi)}=-\frac{\xi}{4\pi c}\sum\limits_{l=1}^{\infty}(2l+1)\bigg\{r_{lTE}[h^{(1)}_l(kr)]^2\\
+r_{lTM}\bigg[l(l+1)\frac{[h^{(1)}_l(kr)]^2}{(kr)^2}+\frac{[krh^{(1)}_l(kr)]'^2}{(kr)^2}\bigg]
\bigg\}
\end{multline}
which together with Eq.~(\ref{CPpotential}) renders the CP potential \eqref{cp_potential_sphere} for a sphere and a ground-state atom.

\end{document}